\documentclass[aps,prb,superscriptaddress,twocolumn,showpacs,amsmath,amssymb]{revtex4}
\usepackage{graphicx}
\usepackage{bm}
\usepackage{color}
\def\be{\begin{equation}}
\def\ee{\end{equation}}
\def\ba{\begin{array}{lll}}
\def\ea{\end{array}}
\def\ber{\begin{eqnarray}}
\def\eer{\end{eqnarray}}

\begin{document}
\title{Plasmons and Coulomb drag in Dirac/Schroedinger hybrid electron systems}
\author{Alessandro Principi}
\affiliation{NEST, Istituto Nanoscienze-CNR and Scuola Normale Superiore, I-56126 Pisa, Italy}
\author{Matteo Carrega}
\affiliation{NEST, Istituto Nanoscienze-CNR and Scuola Normale Superiore, I-56126 Pisa,
Italy}
\author{Reza Asgari}
\email{asgari@ipm.ir}
\affiliation{School of Physics, Institute for Research in Fundamental Sciences (IPM), Tehran 19395-5531, Iran}
\author{Vittorio Pellegrini}
\email{vp@sns.it}
\affiliation{NEST, Istituto Nanoscienze-CNR and Scuola Normale Superiore, I-56126 Pisa,
Italy}
\author{Marco Polini}
\email{m.polini@sns.it}
\affiliation{NEST, Istituto Nanoscienze-CNR and Scuola Normale Superiore, I-56126 Pisa,
Italy}

\date{\today}

\begin{abstract}
We show that the plasmon spectrum of an ordinary two-dimensional electron gas (2DEG) hosted in a GaAs heterostructure is significantly modified when a graphene sheet is placed on the surface of the semiconductor in close proximity to the 2DEG. Long-range Coulomb interactions between massive electrons and massless Dirac fermions lead to a new set of optical and acoustic intra-subband plasmons. Here we compute the dispersion of these coupled modes within the Random Phase Approximation, providing analytical expressions in the long-wavelength limit that shed light on their dependence on the Dirac velocity and Dirac-fermion density. We also evaluate the resistivity in a Coulomb-drag transport setup. These Dirac/Schroedinger hybrid electron systems are experimentally feasible and open new research opportunities for fundamental studies of electron-electron interaction effects in two spatial dimensions.
\end{abstract}

\maketitle

\section{Introduction}
\label{sect:intro}

Plasmons are ubiquitous high-frequency collective density oscillations of an electron liquid, which occur in metals and semiconductors~\cite{Giuliani_and_Vignale}. Their importance across different fields of basic and applied physics is by now well established. They play, for example, a key role in plasmonics~\cite{Maier} and in the photodetection of far-infrared radiation based on field-effect transistors~\cite{knap,vitiello}.

Plasmons in doped graphene sheets (hereafter dubbed ``Dirac plasmons") have been intensively investigated by electron energy-loss spectroscopy (EELS)~\cite{EELS}. 
EELS, however, does not have the sufficient energy and wave-number resolution to address {\it e.g.} the impact of broken Galilean invariance on Dirac plasmons~\cite{abedinpour_prb_2011}. Moreover, its application is limited to the collective modes in the charge channel with no information on the spin degrees of freedom, since the method probes only the ``loss" function, {\it i.e.} the imaginary part of the inverse 
dielectric function $\varepsilon(q,\omega)$. Angle-resolved photoemission spectroscopy~\cite{ARPES} and scanning tunneling spectroscopy~\cite{STM} also give precious information on Dirac plasmons, albeit in a slightly-indirect way.
Finally, Dirac plasmons have been probed by engineering directly their coupling to far-infrared light in a number of intriguing ways~\cite{coupling}. On the contrary, the application of inelastic light (Raman) scattering to probe plasmons in graphene, has been, so far, limited to the investigation of magneto-plasmons in high magnetic fields~\cite{faugeras}.
\begin{figure}[t]
\begin{center}
\includegraphics[width=0.7\linewidth]{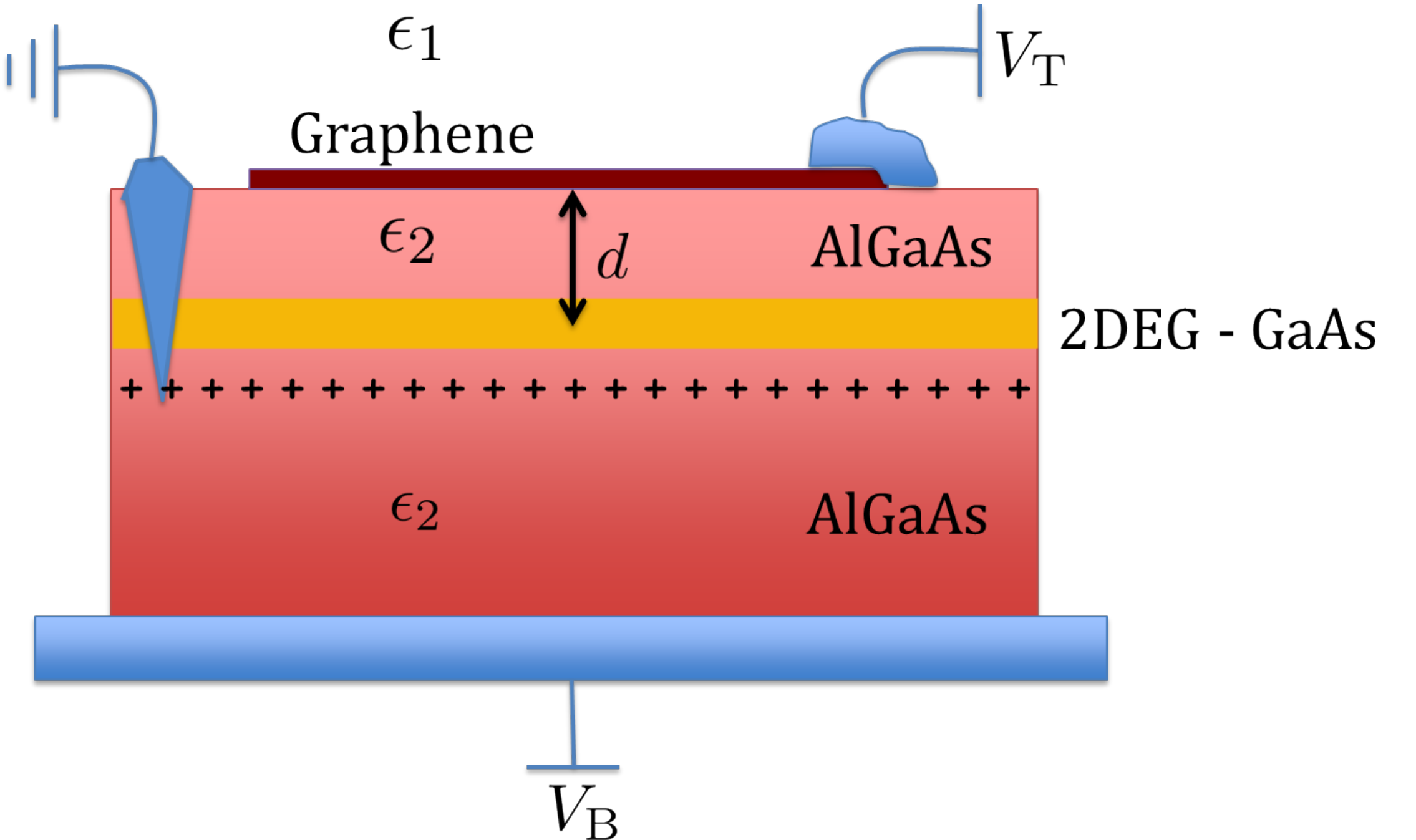}
\caption{(Color online) Schematics of the double-layer massless Dirac/Schroedinger hybrid electron system studied in this work. A graphene sheet is deposited on the surface of a semiconductor, underneath which a GaAs quantum well (at a distance $d$ from the surface) hosts a high-mobility two-dimensional electron gas (2DEG). Carriers in the two layers are induced by conventional gating techniques (the $+$ signs below the quantum well indicate the doping layer). In the presence of Coulomb coupling between the two subsystems, optical and acoustic hybrid collective modes emerge, which can be probed by resonant inelastic light scattering.\label{fig:one}}
\end{center}
\end{figure}

The situation is very different for the case of plasmons in 2DEGs realized, for example, in GaAs/AlGaAs modulation-doped semiconductor heterostructures. Indeed, these excitations have been successfully studied, both in zero magnetic field and in the quantum Hall regime, by resonant inelastic light scattering~\cite{pinczuk,ILS}. This method has a much better energy resolution than EELS (typically below $0.1~{\rm meV}$) and can be straightforwardly generalized to probe spin excitations. The cross section for scattering of photons by the electronic subsystem is, however, tiny. An ingenious resonance condition can be used in a light scattering experiment to enhance this cross section dramatically~\cite{ILS}. The frequency of the incoming or scattered photon can indeed be tuned close to the semiconductor band gap ({\it e.g.} $\approx 1.5~{\rm eV}$ for GaAs at low temperature) making the energy denominators appearing in second-order perturbation theory~\cite{Berestetskii} small and thus yielding a much higher sensitivity to collective electronic modes. 

Graphene~\cite{2007reviews} is a gapless material though, and thus presents immediately an obstacle for the application of resonant inelastic light scattering. The van Hove singularity at the $M$ point of graphene's Brillouin zone can be used to achieve resonance~\cite{chen_nature_2011}, although the band gap at the $M$ point is in the ultraviolet region of the electromagnetic spectrum ($\approx 4~{\rm eV}$) where tunable lasers having the required small linewidth for resonant inelastic light scattering do not exist. 

In this Article we propose a hybrid double-layer system composed by a doped graphene sheet that is Coulomb coupled to an ordinary 2DEG in a GaAs/AlGaAs semiconductor heterostructure, as schematically illustrated in Fig.~\ref{fig:one}. We demonstrate that precious information on Dirac plasmons is encoded in the dispersion of the new collective in-phase and out-of-phase plasmon excitations of this massless Dirac/Schroedinger hybrid system (MDSHS).  These collective modes can in principle be probed {\it via} inelastic light scattering by exploiting, as explained above, the resonance condition offered by the GaAs band gap. In passing, we mention that a similar resonant enhancement mechanism has been recently observed in the case of the Raman phonon lines in epitaxial graphene on copper substrates with incident photon energies overlapping the copper substrate photoluminescence~\cite{costa}. 
We also study transport in a MDSHS in a Coulomb-drag setup~\cite{CD}, demonstrating that, in the low-temperature Fermi-liquid regime and at strong coupling, the drag resistivity displays an intriguing dependence on carrier densities and inter-layer separation, which is different from that known~\cite{CD} for coupled 2DEGs in semiconductor double quantum wells.

Our manuscript is organized as follows. In Sect.~\ref{sect:model} we introduce the model Hamiltonian and some basic definitions. In Sect.~\ref{sect:collectivemodes} we present analytical and numerical results, obtained by means of the Random Phase Approximation, for the collective plasmons modes of MDSHSs. Our main analytical results on the Coulomb drag resistivity are summarized and discussed in Sect.~\ref{sect:coulombdrag}. Finally, in Sect.~\ref{sect:summary} we present a short summary of our main results accompanied by a discussion of the potential of MDSHSs.

\section{Model Hamiltonian and important definitions}
\label{sect:model}

The system depicted in Fig.~\ref{fig:one} can be modeled by the following Hamiltonian:
\begin{eqnarray}\label{eq:fullH}
{\hat {\cal H}} &=& \hbar v_{\rm D}\sum_{{\bm k}, \alpha, \beta} {\hat \psi}^\dagger_{{\bm k}, \alpha} 
( {\bm \sigma}_{\alpha\beta} \cdot {\bm k} ) {\hat \psi}_{{\bm k}, \beta}
+
\frac{1}{2 S}\sum_{{\bm q} \neq {\bm 0}} V_{\rm tt}(q) {\hat \rho}^{({\rm t})}_{\bm q} {\hat \rho}^{({\rm t})}_{-{\bm q}}\nonumber\\
&+& \sum_{{\bm k}} \frac{\hbar^2 {\bm k}^2}{2m_{\rm b}}{\hat \phi}^\dagger_{\bm k}{\hat \phi}_{\bm k} +
\frac{1}{2 S}\sum_{{\bm q} \neq {\bm 0}} V_{\rm bb}(q) {\hat \rho}^{({\rm b})}_{\bm q} {\hat \rho}^{({\rm b})}_{-{\bm q}}\nonumber\\
&+& \frac{1}{2 S}\sum_{{\bm q}} V_{\rm tb}(q) ({\hat \rho}^{({\rm t})}_{\bm q} {\hat \rho}^{({\rm b})}_{-{\bm q}}+{\hat \rho}^{({\rm b})}_{\bm q} {\hat \rho}^{({\rm t})}_{-{\bm q}})~.
\end{eqnarray}
The terms in the first and second lines of the previous equation describe two 2D electron systems with massless Dirac and parabolic Schroedinger bands, respectively. The last term describes Coulomb interactions between the two subsystems. 

In the first line of Eq.~(\ref{eq:fullH}), $v_{\rm D}$ is the Dirac velocity, ${\bm \sigma} = (\sigma^x,\sigma^y)$ is a 2D vector of Pauli matrices, $S$ is the sample area, and ${\hat \psi}^\dagger_{{\bm k}, \alpha}$ (${\hat \psi}_{{\bm k}, \alpha}$) creates (destroys) an electron with momentum ${\bm k}$ and sublattice-pseudospin index $\alpha = A,B$ in graphene.  The Dirac density operator is defined by
$
{\hat \rho}^{({\rm t})}_{\bm q} = \sum_{{\bm k}, \alpha} {\hat \psi}^\dagger_{{\bm k} - {\bm q}, \alpha}{\hat \psi}_{{\bm k}, \alpha}
$. 
In the second line, $m_{\rm b}$ is the bare band mass of the 2DEG ($m_{\rm b} \approx 0.067~m_{\rm e}$ in GaAs, $m_{\rm e}$ being the electron mass in vacuum), ${\hat \phi}^\dagger_{\bm k}$ (${\hat \phi}_{\bm k}$) creates (destroys) an electron with momentum ${\bm k}$ in the 2DEG, and 
$
{\hat \rho}^{({\rm b})}_{\bm q} = \sum_{\bm k} {\hat \phi}^\dagger_{{\bm k} - {\bm q}}{\hat \phi}_{\bm k}
$
is the usual 2DEG density operator. 

In writing Eq.~(\ref{eq:fullH})  we have intentionally hidden ``flavor" indices labeling spin and valley degrees-of-freedom in the graphene part of the Hamiltonian and spin degrees-of-freedom in the 2DEG part of the Hamiltonian. These flavors will just appear as degeneracy factors in the linear response functions $\chi_{\rm t}(q,\omega)$ and $\chi_{\rm b}(q,\omega)$, as we will see below. We have also completely neglected inter-layer tunneling, which is strongly suppressed by the AlGaAs barrier between graphene and the quantum well.

The bare intra- and inter-layer Coulomb interactions, $V_{ij}(q)$ (with $i,j \in \{{\rm t}, {\rm b}\}$), are influenced by the layered
 dielectric environment, modeled by two dielectric constants - $\epsilon_1$ and $\epsilon_2$ in Fig.~\ref{fig:one}. A simple electrostatic calculation~\cite{profumo_prb_2010,formfactor} implies that the Coulomb interaction in the graphene sheet (top layer) is given by
$
V_{\rm tt}(q) = 4\pi e^2/[q (\epsilon_1+\epsilon_2)] 
$,
The Coulomb interaction in the 2DEG (bottom layer) is instead
\begin{equation}\label{eq:v11}
V_{\rm bb}(q) = \frac{4\pi e^2}{q D(q)} [ (\epsilon_2 + \epsilon_1) e^{qd} + 
 (\epsilon_2 - \epsilon_1) e^{-qd}]~,
\end{equation}
where $D(q) =  2\epsilon_2(\epsilon_1 + \epsilon_2) e^{qd}$. Finally, the inter-layer interaction is given by
\begin{equation}\label{eq:v12} 
V_{\rm tb}(q) = V_{\rm bt}(q) = \frac{8\pi e^2}{q D(q)}~\epsilon_2~.
\end{equation}
Note that in the ``uniform" $\epsilon_1 = \epsilon_2 \equiv \epsilon$ limit we recover the familiar expressions 
$V_{\rm tt}(q) = V_{\rm bb}(q) \to 2\pi e^2/(\epsilon q)$ and $V_{\rm tb}(q) = V_{\rm bt}(q) = V_{\rm tt}(q)\exp(-qd)$. 

For the following analysis we introduce the electron density in the top (bottom) layer $n_{\rm t}$ ($n_{\rm b}$) and the corresponding Fermi wave numbers 
$k_{{\rm F}, i} = \sqrt{4 \pi n_i/N_i}$, where $N_i$ is a degeneracy factor ($N_{\rm t} =4$ in the graphene layer and $N_{\rm b} =2$ in the 2DEG layer). We also introduce the Fermi energies $\varepsilon_{{\rm F}, {\rm t}} = \hbar v_{\rm D} k_{{\rm F}, {\rm t}}$ and $\varepsilon_{{\rm F}, {\rm b}} = \hbar^2 k^2_{{\rm F}, {\rm b}}/(2 m_{\rm b})$ in the top and bottom layers, respectively, $\alpha_{\rm ee} = e^2/(\hbar v_{\rm D}) \approx 2.2$,  and $r_s = (\pi n_{\rm b} a^2_{\rm B})^{-1/2}$, $a_{\rm B} = \hbar^2/(m_{\rm b} e^2)$ being the Bohr radius calculated with the semiconductor band mass $m_{\rm b}$ (note the absence of any dielectric constant in the definition of $a_{\rm B}$).

\section{Collective optical and acoustic plasmon modes}
\label{sect:collectivemodes}

The collective modes of the system described by the Hamiltonian (\ref{eq:fullH}) can be 
determined~\cite{Giuliani_and_Vignale} by locating the poles of the linear-response function ${\bm \chi}(q,\omega)$. 
Within the Random Phase Approximation (RPA) these functions satisfy the following matrix equation~\cite{Giuliani_and_Vignale}:
\begin{equation}\label{eq:matrix-form}
{\bm \chi}^{-1}(q,\omega) = {\bm \chi}^{-1}_0(q,\omega) - {\bm V}(q)~,
\end{equation}
where ${\bm \chi}_0(q,\omega)$ is a $2 \times 2$ diagonal matrix whose elements $\chi_{\rm t}(q,\omega)$ and $\chi_{\rm b}(q,\omega)$ are the well-known non-interacting (Lindhard) response functions of each layer at arbitrary doping $n_i$.  The mathematical and physical properties of $\chi_{\rm t}(q,\omega)$ are discussed in Ref.~\onlinecite{lindhardMDF}, while expressions for $\Re e~[\chi_{\rm b}(q,\omega)]$ and $\Im m~[\chi_{\rm b}(q,\omega)]$ can be found in Ref.~\onlinecite{Giuliani_and_Vignale}. The off-diagonal (diagonal) elements of the matrix ${\bm V} =\{V_{ij}\}_{i,j = {\rm t}, {\rm b}}$ represent inter-layer (intra-layer) Coulomb interactions. 

A straightforward inversion of Eq.~(\ref{eq:matrix-form}) yields the following condition for collective modes:
\begin{eqnarray}\label{eq:zeroes}
\varepsilon(q,\omega) &=& [1-V_{\rm tt}(q) \chi_{\rm t}(q,\omega)][1-V_{\rm bb}(q)\chi_{\rm b}(q,\omega)] \nonumber\\
&-& V^2_{\rm tb}(q)\chi_{\rm t}(q,\omega)\chi_{\rm b}(q,\omega) =0~.
\end{eqnarray}
Zeroes of $\varepsilon(q,\omega)$ occur above the intra-band particle-hole 
continuum where $\chi_i$ is real, positive, and a decreasing function of 
frequency.  Eq.~(\ref{eq:zeroes}) admits two solutions~\cite{dassarma_prb_1981}, a higher frequency 
solution at $\omega_{\rm op}(q \to 0) \propto \sqrt{q}$ which corresponds to in-phase oscillations of the 
densities in the two layers (optical plasmon), and a lower frequency solution at 
$\omega_{\rm ac}(q \to 0) \propto q$ which corresponds to out-of-phase oscillations (acoustic plasmon).

\begin{figure}[t]
\begin{center}
\includegraphics[width=1.00\linewidth]{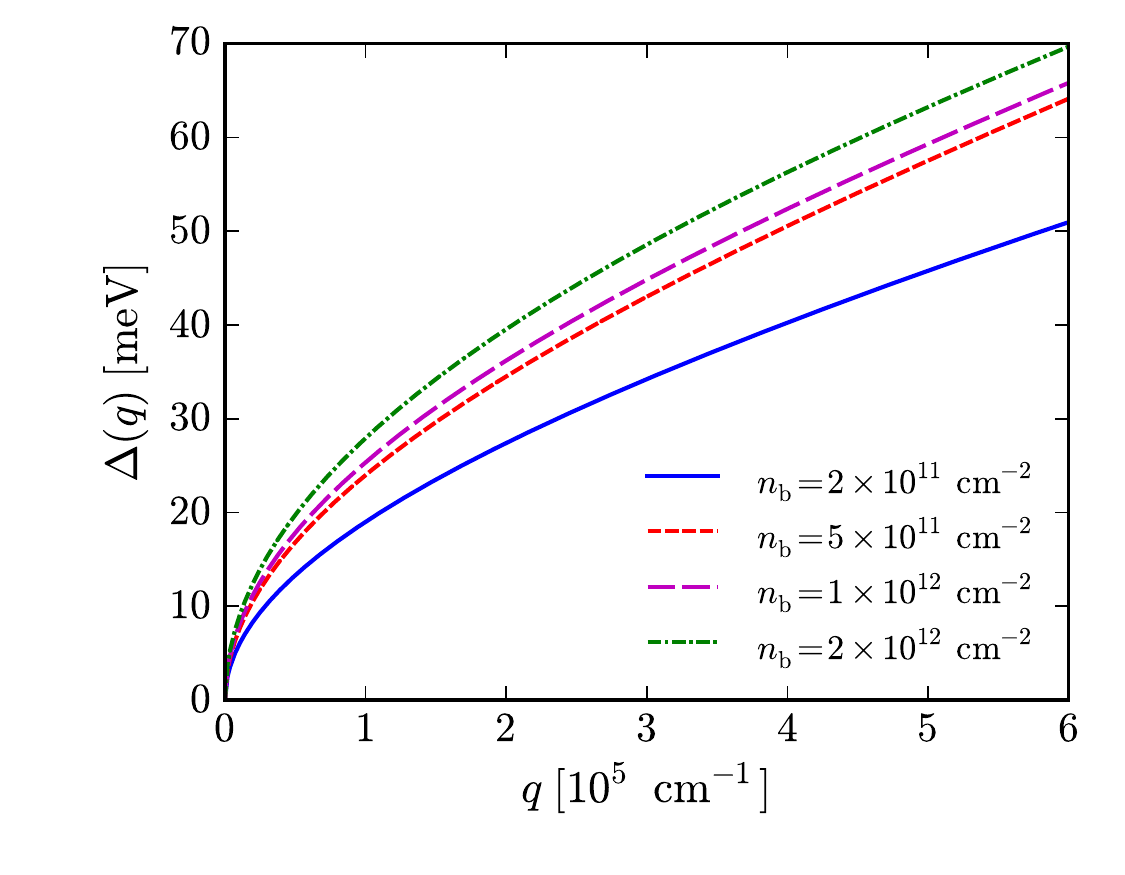}
\caption{(Color online) Optical plasmon dispersion in a massless Dirac/Schroedinger hybrid electron system. In this figure we plot the quantity $\Delta(q) = \hbar\omega_{\rm op}(q) - \hbar\omega_{\rm pl}(q)$ (in ${\rm meV}$) as a function of the wave number $q$ (in units of $10^5~{\rm cm}^{-1}$). Different line styles refer to different values of the 2DEG carrier density $n_{\rm b}$, ranging from $n_{\rm b}=1\times 10^{11}~{\rm cm}^{-2}$ to $n_{\rm b} = 1\times 10^{12}~{\rm cm}^{-2}$. The Dirac-fermion density is fixed at the value $n_{\rm t} = 10^{12}~{\rm cm}^{-2}$, while the inter-layer distance is $d = 30~{\rm nm}$. With reference to Fig.~\ref{fig:one}, the dielectric constants have been fixed to $\epsilon_1=1$ and $\epsilon_2=13$. Note that $\Delta(q)$ is positive, implying a blue shift of the optical plasmon mode of the hybrid system with respect to the usual plasmon mode of an isolated 2DEG.\label{fig:two}}
\end{center}
\end{figure}

\begin{figure}[t]
\begin{center}
\includegraphics[width=1.00\linewidth]{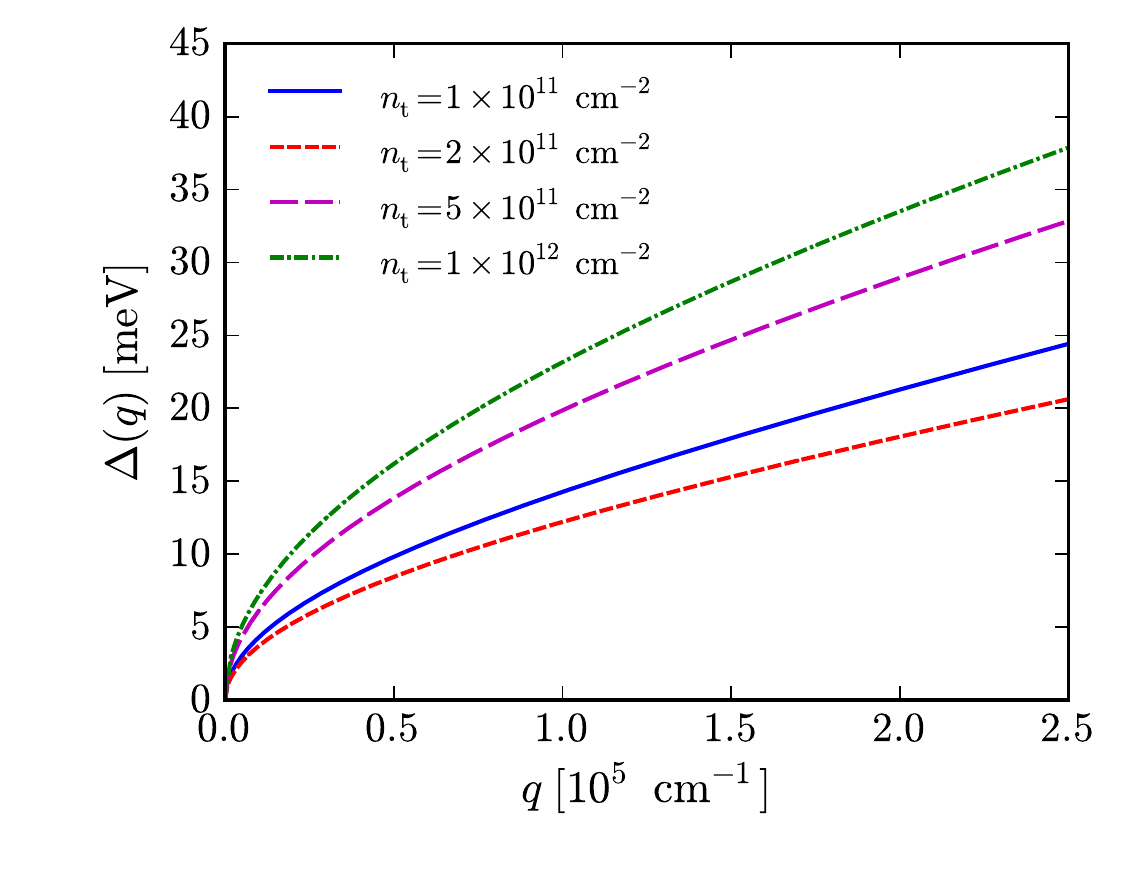}
\caption{(Color online) Same as in Fig.~\ref{fig:two} but for different values of the Dirac-fermion density $n_{\rm t}$ and for a fixed value of the 2DEG carrier density $n_{\rm b} = 2\times 10^{11}~{\rm cm}^{-2}$. Notice the non-monotonic behavior of $\Delta(q)$: for each $q$, $\Delta(q)$ reaches its minimum when the Dirac-fermion density matches the 2DEG density ($n_{\rm t}=n_{\rm b}$).\label{fig:three}}
\end{center}
\end{figure}

A summary of our main results for $\omega_{\rm op}(q)$, obtained from the numerical solution of Eq.~(\ref{eq:zeroes}), is reported in Figs.~\ref{fig:two}-\ref{fig:three}. In these figures we plot the quantity $\Delta(q) = \hbar\omega_{\rm op}(q) - \hbar \omega_{\rm pl}(q)$, which physically represents the energy of the MDSHS optical plasmon measured from that of the plasmon of an {\it isolated} 2DEG, $\hbar\omega_{\rm pl}(q)$. We clearly see that $\Delta(q)$ is positive and $\approx 10~{\rm meV}$ at $q \sim 10^5~{\rm cm}^{-1}$. 

To leading order in $q$ in the long-wavelength $q \to 0$ limit, the frequency of the MDSHS optical plasmon mode can be found analytically. The result is
\be\label{eq:opticalplasmon}
\omega^2_{\rm op}(q\to 0) = \left(\frac{N_{\rm t} e^2 v_{\rm D} k_{\rm F, t}}{2 \hbar {\bar \epsilon}} + \frac{2\pi n_{\rm b} e^2}{m_{\rm b} {\bar \epsilon}}\right) q~,
\ee
where we have introduced ${\bar \epsilon} \equiv (\epsilon_1 +\epsilon_2)/2$.
We have checked that Eq.~(\ref{eq:opticalplasmon}) is in perfect agreement with the numerical results displayed in Figs.~\ref{fig:two}-\ref{fig:three}. The first term on the r.h.s. of Eq.~(\ref{eq:opticalplasmon}) can be easily recognized~\cite{lindhardMDF} to be the square of the RPA plasmon frequency of the electron gas in an isolated graphene sheet separating two media with dielectric constants $\epsilon_1$ and $\epsilon_2$. The second term is the well-known RPA plasmon frequency of a 2D parabolic-band electron gas~\cite{Giuliani_and_Vignale}. A measurement of $\omega_{\rm op}(q)$ in a MDSHS at sufficiently small $q$ [more precisely, for $q \ll \min(k_{\rm F,t}, k_{\rm F, b})$] thus allows to access directly the Dirac velocity $v_{\rm D}$ and the Dirac-fermion density $n_{\rm t} \propto k^2_{\rm F, t}$.

\begin{figure}[t]
\begin{center}
\includegraphics[width=1.00\linewidth]{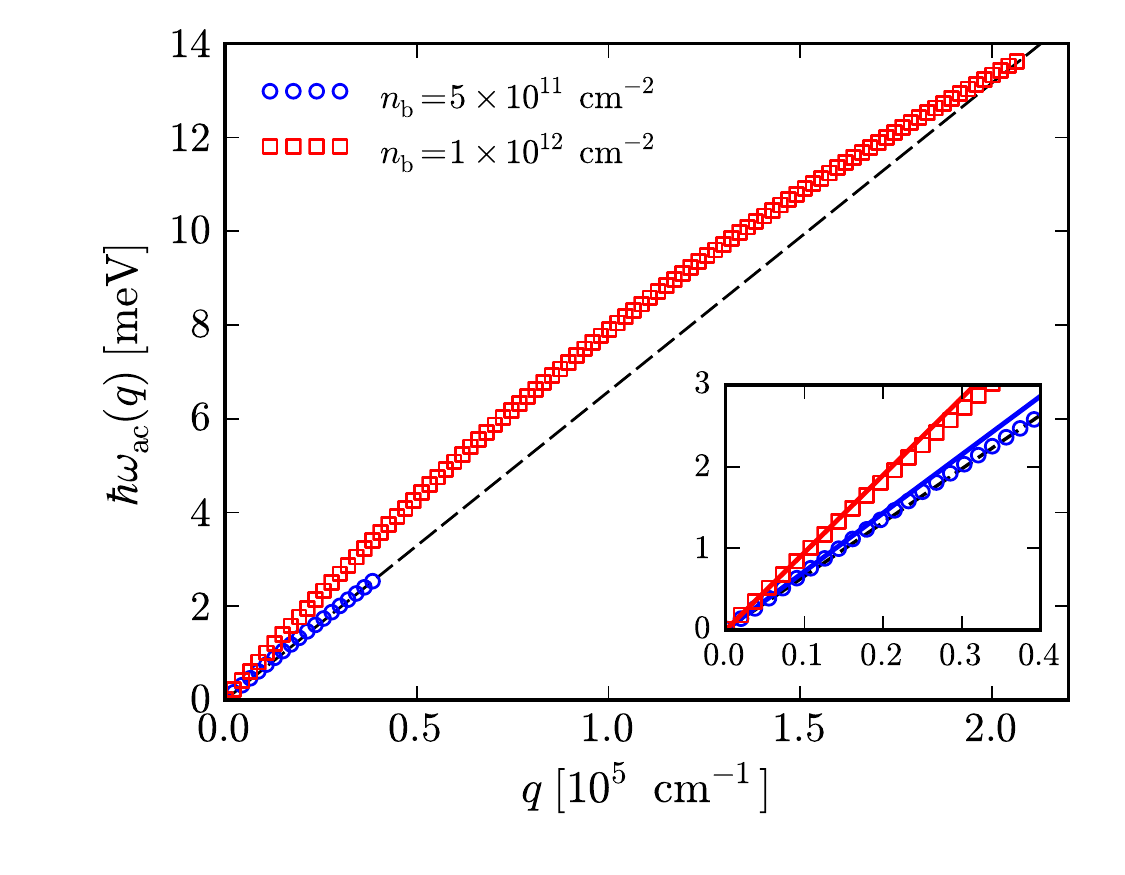}
\caption{(Color online) The acoustic plasmon mode of a massless Dirac/Schroedinger hybrid system. 
The acoustic plasmon dispersion $\hbar\omega_{\rm ac}(q)$ (in ${\rm meV}$) 
is plotted as a function of the wave number $q$ (in units of $10^5~{\rm cm}^{-1}$). Empty symbols label data for two 
different values of the 2DEG carrier density $n_{\rm b}$. In this figure we have fixed 
$n_{\rm t} = 1 \times 10^{11}~{\rm cm}^{-2}$ and $d = 60~{\rm nm}$. The other parameters are identical to those used in Figs.~\ref{fig:two}-\ref{fig:three}. In the inset we illustrate the comparison between numerical results for the acoustic plasmon dispersion (empty symbols) and the analytical result $\omega_{\rm ac}(q\to 0) = c_{\rm s} q$ (solid lines) with the velocity $c_{\rm s}$ obtained from the solution of Eq.~(\ref{eq:csound}). The dashed line marks the upper boundary of the Dirac fermion intra-band single particle continuum. \label{fig:four}}
\end{center}
\end{figure}

In Fig.~\ref{fig:four} we show illustrative results for the acoustic plasmon $\omega_{\rm ac}(q)$ as obtained from the numerical solution of Eq.~(\ref{eq:zeroes}) for different values of the 2DEG carrier densities $n_{\rm b}$. The dashed line in this figure represents the upper bound $\omega = v_{\rm D} q$ of the Dirac-fermion intra-band electron-hole continuum~\cite{lindhardMDF}. Acoustic plasmon dispersions lying above the dashed line are not damped (within RPA) since they cannot decay by exciting intra-band electron-hole pairs in the graphene sheet. When the dispersion hits the dashed line, Landau damping occurs and the acoustic plasmon pole acquires a finite lifetime. Note that when the 2DEG density is low (empty circles in Fig.~\ref{fig:four}) the acoustic-plasmon dispersion stays out the Dirac-fermion intra-band particle-hole continuum for a very small range of wave numbers $q$. It will be thus difficult to observe acoustic plasmons in MDSHSs in which the 2DEG carrier density is low.

The acoustic-plasmon group velocity $c_{\rm s} = \lim_{q \to 0} \omega_{\rm ac}(q)/q$ can be calculated analytically following the procedure explained in Ref.~\onlinecite{santoro_prb_1988}.
After some tedious but straightforward algebra we find the following equation for $x = c_{\rm s}/v_{\rm D}$, the ratio between the acoustic-plasmon group velocity $c_{\rm s}$ 
and the Dirac velocity $v_{\rm D}$:
\begin{eqnarray}\label{eq:csound}
&& 
2 N_{\rm t} \alpha_{\rm ee} \xi \Gamma f(x) g(x) -\epsilon_2 \Gamma g(x) \sqrt{x^2-1}
\nonumber\\
&&
-2 \epsilon_2 N_{\rm t} \alpha_{\rm ee}f(x) \sqrt{x^2 N_{\rm b} r_s^2 - 4 \alpha_{\rm ee}^2}=0
~,
\end{eqnarray}
where $\xi = d k_{{\rm F}, {\rm t}}$, $\Gamma = 2N_{\rm b}/(k_{{\rm F}, {\rm t}} a_{\rm B})$, $f(x) = x - \sqrt{x^2-1}$, and $g(x) = x \sqrt{N_{\rm b}} r_s - \sqrt{x^2 N_{\rm b} r_s^2 - 4 \alpha_{\rm ee}^2}$. Note that the solution of Eq.~(\ref{eq:csound}) depends only on $\epsilon_2$, while, as we have seen above, $\omega_{\rm op}(q\to 0)$ depends on the average ${\bar \epsilon}$.  

An undamped acoustic plasmon emerges for $x>x_{\rm crit}$, where the threshold $x_{\rm crit}$ can be directly found from Eq.~(\ref{eq:csound}). When the Dirac velocity $v_{\rm D}$ is larger [smaller] than the Fermi velocity $v_{{\rm F}, {\rm b}} =\hbar k_{\rm F, b}/m_{\rm b}$ in the 2DEG, $x_{\rm crit}=1$ [$x_{\rm crit}=v_{{\rm F},{\rm b}}/v_{\rm D} = 2\alpha_{\rm ee}/(r_s \sqrt{N_{\rm b}})$]. 
The existence of a solution of Eq.~(\ref{eq:csound}) thus depends on three parameters, namely $d$, $n_{\rm t}$, and $n_{\rm b}$ (when the dielectric constants $\epsilon_1$ and $\epsilon_2$ are fixed). For example, given $n_{\rm t}$ and $n_{\rm b}$, the acoustic plasmon emerges out of the continuum for $d>d^{({\rm crit})}$ with
\be
d^{({\rm crit})} = \left\{
\begin{array}{ll} 
{\displaystyle \frac{\epsilon_2 \sqrt{N_{\rm b} r_s^2 - 4 \alpha_{\rm ee}^2}}
{\Gamma k_{{\rm F},{\rm t}} g(1)} }, & {\rm if}~v_{\rm D}>v_{{\rm F},{\rm b}}
\vspace{0.2cm}\\
{\displaystyle \frac{\epsilon_2 v_{{\rm F},{\rm b}} 
\sqrt{4 \alpha_{\rm ee}^2 -N_{\rm b} r_s^2}}{4 N_{\rm t} 
\alpha_{\rm ee}^2 v_{\rm D} k_{{\rm F},{\rm t}} f(v_{{\rm F},{\rm b}}/v_{\rm D})} }~, & {\rm otherwise}
\end{array}
\right.~.
\ee
We remind the reader that the functions $f(x)$ and $g(x)$ have been defined right after Eq.~(\ref{eq:csound}). 
Similarly, given $n_{\rm t}$ and $d$, the acoustic plasmon emerges out of the continuum 
for $n>n_{\rm b}^{({\rm crit})}$, where
\be	
n_{\rm b}^{({\rm crit})} = \left\{
\begin{array}{ll}
{\displaystyle \frac{\epsilon_2 N_{\rm b} (\epsilon_2 + 2 \xi \Gamma)}
{4 \pi^2 a_{\rm B}^2 \alpha_{\rm ee}^2 (\epsilon_2 + \xi \Gamma)^2}}~, & {\rm if}~v_{\rm D}>v_{{\rm F},{\rm b}}
\vspace{0.2cm}\\
{\displaystyle \frac{N_{\rm b} (\epsilon_2 + 2 \xi N_{\rm t} \alpha_{\rm ee})^2}
{4 \pi \epsilon_2 a_{\rm B}^2 \alpha_{\rm ee}^2 (\epsilon_2 + 4 \xi N_{\rm t} \alpha_{\rm ee})} }~, & {\rm otherwise}
\end{array}
\right.~.
\ee
Note that the critical values of $d$ and $n_{\rm b}$ do not depend on $n_{\rm t}$ when $v_{\rm D}>v_{{\rm F},{\rm b}}$. 
They indeed depend only on the products $\Gamma k_{{\rm F},{\rm t}}$ and $\xi \Gamma$, 
which are independent of $k_{\rm F,t}$.

The theoretical prediction based on the solution of Eq.~(\ref{eq:csound}) is compared with the numerical results in the inset to Fig.~\ref{fig:four}. 
\section{Coulomb drag resistivity}
\label{sect:coulombdrag}

In a drag experiment~\cite{CD} a constant current is imposed on one layer  (the ``active" or ``drive" layer). If no current is allowed to flow in the other one (the ``passive" layer),  an electric field develops, whose associated force cancels the frictional drag force exerted by the electrons in the active layer on the electrons in the passive one. The drag resistivity $\rho_{\rm D}$, is defined as the ratio of the induced voltage in the passive layer to the applied current in the drive layer and reads $\rho_{{\rm D}} = - \sigma_{\rm D}/(\sigma_{\rm t} \sigma_{\rm b} - \sigma^2_{\rm D}) \simeq - \sigma_{\rm D}/(\sigma_{\rm t} \sigma_{\rm b})$ in terms of the intra-layer conductivities $\sigma_{{\rm t}}$ ($\sigma_{{\rm b}}$) of the top (bottom) layer and the drag conductivity $\sigma_{\rm D}$.
Notice that in writing the last equality of the previous equation we have assumed $\sigma_{\rm D} \ll \sigma_{\rm t}, \sigma_{\rm b}$.

Within the Kubo formalism the drag conductivity at the second order in the screened inter-layer interaction $U_{\rm tb}(q,\omega)$ reads~\cite{kamenev_prb_1995} ($\hbar = 1$)
\be\label{eq:conductivity_end}
\sigma_{\rm D} = \frac{\beta e^2}{8\pi} \int\! \frac{d^2{\bm q}}{(2\pi)^2}
\! \int_{0}^{\infty} d\omega \frac{|U_{{\rm tb}}(q,\omega)|^2}{\sinh^2(\beta\omega/2)}
\Gamma_{\rm t}({\bm q},\omega) \Gamma_{\rm b}({\bm q},\omega)~,
\ee
where $\beta = (k_{\rm B} T)^{-1}$ and $\Gamma_i({\bm q},\omega)$ is the so-called ``non-linear susceptibility" of the $i$-th layer.
The dynamically-screened inter-layer interaction that appears in Eq.~(\ref{eq:conductivity_end}) is 
$U_{\rm tb}(q,\omega) = V_{\rm tb}(q)/\varepsilon (q,\omega)$, 
where $\varepsilon (q,\omega)$ is the RPA dielectric function defined in Eq.~(\ref{eq:zeroes}).

Here we focus only on the low-temperature limit, {\it i.e.} $k_{{\rm B}} T \ll \min (\varepsilon_{{\rm F},{\rm t}}, \varepsilon_{{\rm F},{\rm b}})$.
In this regime we can do the following approximations~\cite{carrega_arxiv_2012}: (i) use the low-temperature expressions for the intra-layer conductivities $\sigma_i$, (ii) substitute $U_{{\rm tb}}(q,\omega)$ in Eq.~(\ref{eq:conductivity_end}) with the statically-screened inter-layer interaction $U_{{\rm tb}}(q, \omega=0)$, and (iii) expand the non-linear susceptibilities $\Gamma_i({\bm q} , \omega \to 0)$ up to the lowest order in $\omega$.

Following Ref.~\onlinecite{carrega_arxiv_2012}, we find the following results for the top layer (graphene sheet):
\be
\label{eq:sigma_1_lowt}
\lim_{T \to 0} \sigma_{\rm t} = \frac{e^2}{4\pi} N_{\rm t} \tau_{\rm t} (k_{\rm F, t})\varepsilon_{\rm F, t}
\ee
and
\ber\label{eq:gamma1lowt}
\lim_{T \to 0} \Gamma_{{\rm t}} ({\bm q}, \omega\to 0) &=& - N_{\rm t}\frac{\omega \tau_{\rm t}(k_{{\rm F, t}})}{2\pi v_{\rm D}}\Theta(2k_{{\rm F, t}} -q)  \nonumber \\
&\times&\left(1- \frac{q^2}{4 k^2_{{\rm F, t}}}\right)^{1/2} \cos(\varphi_{{\bm q}})~,
\eer
and the following results for the bottom layer (2DEG):
\be
\label{eq:sigma_2_lowt}
\lim_{T \to 0} \sigma_{\rm b} = \frac{e^2}{2\pi} N_{\rm b} \tau_{\rm b}(k_{\rm F, b}) \varepsilon_{{\rm F, b}} = \frac{n_{\rm b} e^2 \tau_{\rm b} (k_{\rm F, b})}{m_{\rm b}}
\ee
and
\ber
\label{eq:gamma2lowt}
\lim _{T \to 0} \Gamma_{\rm b} ({\bm q} , \omega \to 0) &=& - N_{\rm b}\frac{ \omega \tau_{\rm b} (k_{{\rm F, b}})}{4\pi v_{\rm F, b}} \Theta ( 2 k_{{\rm F, b}} - q)
\nonumber \\
&\times & \left(1 - \frac{q^2}{4 k^2_{{\rm F, b}}}\right)^{-1/2}\cos (\varphi_{{\bm q}}) ~,
\eer
where $v_{{\rm F}, {\rm b}}$ is the 2DEG Fermi velocity. Note the singularity at $q= 2k_{\rm F,b}$ in the non-linear susceptibility in Eq.~(\ref{eq:gamma2lowt}). On the contrary, $\Gamma_{{\rm t}} ({\bm q}, \omega\to 0)$ vanishes at $q= 2k_{\rm F,t}$, as a consequence of the absence of backscattering for massless Dirac fermions~\cite{2007reviews}.

Using Eqs.~(\ref{eq:sigma_1_lowt})-(\ref{eq:gamma2lowt}) we find that the drag resistivity in the low-temperature limit is given by
\ber
\label{eq:rho_hybridT0}
\lim_{T\to 0}\rho_{\rm D} &=& - \frac{1 }{24 e^2} \frac{1}{v_{\rm D} v_{{\rm F, b}}}\frac{(k_{{\rm B}}T)^2}{ \varepsilon_{{\rm F, t}}\varepsilon_{{\rm F, b}}}  \int_0 ^{q_{{\rm max}}} q ~ d q | U_{{\rm tb} }(q,0)|^2
\nonumber \\
&\times& {\cal F}\left(\frac{q}{2 k_{\rm F, t}}, \frac{q}{2 k_{\rm F, b}}\right)~,
\eer
where $q_{{\rm max}} = \min( 2 k_{{\rm F, t}} , 2k_{{\rm F, b}})$ and ${\cal F}(x,y) = \sqrt{(1-x^2)/(1-y^2)}$. 

It is convenient at this stage to introduce a dimensionless expansion parameter $\eta \equiv d \sqrt{k_{\rm F, t}k_{\rm F, b}}$. The MDSHS is weakly (strongly) coupled if $\eta \gg 1$ ($\eta \ll 1$). Similar reasoning to the one employed in Ref.~\onlinecite{carrega_arxiv_2012} yields the following result for $\rho_{\rm D}$ in the weak-coupling limit (restoring Planck's constant for clarity):
\ber\label{eq:weak_1}
\lim_{\eta \to \infty}\lim_{T \to 0} \rho_{\rm D} &=& - \frac{h }{ e^2}\frac{\pi \zeta (3) \epsilon^2_2}{16 \alpha^2_{\rm ee} N^2_{\rm t} N^2_{\rm b}} \frac{(k_{\rm B}T)^2}{ \hbar^2 v^2_{\rm D} k^3_{\rm F, t} k^3_{\rm F, b} d^4} 
\nonumber\\
&\propto& - \frac{h}{e^2}\frac{T^2}{ n_{\rm t}^{3/2} n_{\rm b}^{3/2}d^4}~.
\eer
We emphasize that the functional dependence of $\rho_{\rm D}$ on the carrier densities $n_i$ and on the inter-layer distance $d$ in this limit 
is the same as in the case of two 2DEGs~\cite{flensberg_prb_1995} and two graphene sheets~\cite{tse_prb_2007,carrega_arxiv_2012}.

In the strong-coupling limit, we find
\ber \label{eq:strong_1}
\lim_{\eta \to 0}\lim_{T \to 0}\rho_{{\rm D}} &=& - \frac{h }{ e^2}\frac{\pi}{3} \alpha^2_{\rm ee}\frac{v_{\rm D}}{ v_{\rm F, b}}\frac{(k_{{\rm B}}T)^2}{ \varepsilon_{{\rm F, t}}\varepsilon_{{\rm F, b}}}
\int_0 ^{x_{{\rm max}}}  d x ~ x
\nonumber\\
&\times&
\frac{\displaystyle {\cal F}\left(\sqrt{\frac{k_{\rm F, b}}{k_{\rm F, t}}}\frac{x}{2}, \sqrt{\frac{k_{\rm F, t}}{k_{\rm F, b}}}\frac{x}{2} \right)}{\left[2 {\bar \epsilon} x + 2(q_{\rm TF, t} + q_{\rm TF, b})/\sqrt{k_{\rm F, t}k_{\rm F, b}}~\right]^2}~,\nonumber\\
\eer
where $x_{\rm max} = \min( 2\sqrt{k_{\rm F,t}/k_{\rm F,b}}, 2 \sqrt{k_{\rm F,b}/k_{\rm F, t}})$ and we have introduced the Thomas-Fermi screening wave numbers $q_{\rm TF, t} = N_{\rm t}\alpha_{\rm ee} k_{\rm F, t}$ and $q_{\rm TF, b}= N^{3/2}_{\rm b} r_s k_{\rm F, b}/2$.

The quadrature on the r.h.s. of Eq.~(\ref{eq:strong_1}) can be carried out analytically in the special case $k_{\rm F, t} = k_{\rm F, b}$. Due to the different degeneracies in two layers, this condition implies a density imbalance between the two layers: $n_{\rm b} = N_{\rm b} n_{\rm t}/N_{\rm t} \equiv n$. In this case, Eq.~(\ref{eq:strong_1}) yields
\ber\label{eq:strongcouplingsimplified}
\lim_{\eta \to 0}\lim_{T \to 0}\rho_{\rm D}\big|_{k_{\rm F, t} = k_{\rm F, b}} &=& - \frac{h }{e^2}\frac{\pi}{12}\frac{v_{\rm D}}{ v_{\rm F, b}} \frac{(k_{\rm B}T)^2}{\varepsilon_{\rm F, b}\varepsilon_{\rm F, t}}\frac{\alpha_{\rm ee}^2}{{\bar \epsilon}^2}
\nonumber\\
&\times&
\Bigg\{
\ln \left[1 + \frac{2 {\bar \epsilon}}{N_{\rm t} \alpha_{\rm ee} + N^{3/2}_{\rm b} r_s/2}\right] 
\nonumber\\
&-& \frac{2{\bar \epsilon}}{2 {\bar \epsilon}+ N_{\rm t} \alpha_{\rm ee} + N^{3/2}_{\rm b}r_s/2}
\Bigg\}
~.
\nonumber\\
\eer
Eqs.~(\ref{eq:rho_hybridT0})-(\ref{eq:strongcouplingsimplified}) are the most important results of this Section~\cite{scharf_arxiv_2012}. 
As in the case of drag between two graphene sheets~\cite{carrega_arxiv_2012}, 
Eq.~(\ref{eq:strongcouplingsimplified}) does not depend on the inter-layer distance $d$. In the limit $n\to 0$ it yields a dependence of $\rho_{\rm D}$ on carrier density of the form $\rho_{\rm D} \propto n^{-1}$, analogously to what found by Carrega {\it et al.}~\cite{carrega_arxiv_2012} for two graphene sheets with equal density [note that in the limit $n \to 0$ one has to find the asymptotic behavior of the expression in curly brackets in Eq.~(\ref{eq:strongcouplingsimplified}) for $r_s \to \infty$]. Finally, we emphasize that Eq.~(\ref{eq:strongcouplingsimplified}) does not contain the well-known~\cite{CD} $\ln(T)$ factor which appears in the drag resistivity at strong coupling in the case of two 2DEGs in semiconductor double quantum wells. This factor, which stems from the contribution to drag from momenta $q$ of the order of  $2k_{\rm F}$, is completely suppressed in the present case by the absence of backscattering for massless Dirac fermions~\cite{2007reviews} [compare Eq.~(\ref{eq:gamma1lowt}) with Eq.~(\ref{eq:gamma2lowt})].

\section{Summary and discussion}
\label{sect:summary}

In this work we have proposed that massless Dirac/Schroedinger hybrid double layers be used to probe Dirac plasmons in a graphene sheet by employing resonant inelastic light scattering~\cite{ILS}. A natural resonant condition in this system is offered by the band gap of the semiconductor ({\it e.g.} GaAs) hosting an ordinary parabolic-band 2D electron gas, thus bypassing the non-trivial issue of the absence of a gap in an isolated graphene sheet.  We have demonstrated that information on Dirac plasmons can be extracted in a rather direct manner from the measurement of the optical plasmon mode of the hybrid double layer. The latter supports also a soft mode which disperses linearly as a function of wave number and whose observation requires a sufficiently-high electron density in the semiconductor quantum well. Finally, we have calculated the low-temperature Coulomb-drag resistivity in the Boltzmann-transport limit, showing that in the strong-coupling limit it displays a dependence on carrier densities that is not shared by conventional all-semiconductor double quantum wells~\cite{CD}. 

The system depicted in Fig.~\ref{fig:one} is amenable to experimental investigations and paves the way for the study of many other intriguing phenomena. We expect, for example, that massless Dirac/Schroedinger hybrid double layers will display interesting correlated states (and associated transport anomalies) induced by inter-layer interactions deep in the quantum Hall regime. Another very appealing subject of investigation could be ``hybrid exciton condensates". Exciton condensates are elusive many-particle systems in which electron-hole pairs condense below a certain critical temperature in a coherent superfluid~\cite{ECs}. Double-layer structures are extremely useful to spatially separate electrons and holes in two different layers, thereby suppressing electron-hole recombination~\cite{lozovik_JEPT_1975}. Realizing all-semiconductor-based electron-hole double layers is a rather difficult task~\cite{electronholesemiconductors}. On the contrary, holes can be trivially induced in the graphene sheet depicted in Fig.~\ref{fig:one} by gating techniques. Massless Dirac/Schroedinger hybrid double layers thus offer unprecedented opportunities to realize, probe, and manipulate novel electron-hole quantum liquids and, hopefully, hybrid exciton condensates at sufficiently low temperatures.

\acknowledgements 

Work in Pisa was supported by MIUR through the program ``FIRB - Futuro in Ricerca 2010", Grant no. RBFR10M5BT (``Plasmons and terahertz devices in graphene"). We thank Allan MacDonald, Leonid Levitov, Vincenzo Piazza, Aron Pinczuk, and Giovanni Vignale for useful discussions.


\begin{thebibliography}{77}
\bibitem{formfactor}
	A form factor related to the finite width of the quantum well where the 2DEG is hosted 
	has been neglected for the sake of simplicity and has the effect of weakening $V_{\rm bb}(q)$. 
	Its introduction does not pose any conceptual difficulty and does not change our results qualitatively.
\end{thebibliography}
\end{document}